# THERMODYNAMICALLY PREDICTED CHEMICAL OSCILLATIONS IN CLOSED CHEMICAL SYSTEMS

B. Zilbergleyt[I]

INTRODUCTION

All known up to now models of chemical oscillations are based exclusively on kinetic considerations. The chemical gross-process equation is split usually by elementary steps, each step is supplied by an arrow and a differential equation. Joint solution to such a construction leads to chemical oscillations under certain, often *ad hoc* chosen conditions and with *ad hoc* numerical coefficients: they predict chemical oscillations mostly within narrow range of parameters. A series of once upon a time revolutionary models like "brusselator" [1], "oregonator" [2], etc. have been originated this way. Title of one publication – "Chemical oscillations arise solely from kinetic nonlinearity … " [3] – may serve as a manifesto for such an approach. Kinetic perception of chemical oscillations reigns in numerous publications and basic monographs [4], including some recent [5,6]. Also, chemical instabilities as a reason of chemical oscillations are mentioned usually in the most kinetic oscillatory models.

In this work we show that chemical oscillations also may follow from thermodynamic considerations; at the same time we do not deny the ability of kinetic models to produce oscillations, close to observed experimentally. Although chemical oscillations of non-kinetic origin have been predicted by discrete thermodynamics (DTD) in electrochemical [7] and laser systems [8], a general approach to the problem was not yet studied, and this is the topic of the paper.

CHEMICAL SYSTEM MAP OF STATES

From the DTD notion of chemical equilibrium as a state, where thermodynamic forces (TdF), acting against the chemical system–internal (thermodynamic affinity) and external–are mutually balanced, follows the chemical system map of states, e.g for isothermobaric conditions [9]

(1) $\qquad \ln[\Pi_j(\eta_j,0)/\Pi_j(\eta_j,\delta_j)] - \tau_j \varphi(\delta_j,\pi_j) = 0$,

where $\Pi_j(\eta_j,0)$ is mole fraction product at "true" thermodynamic equilibrium (TdE), defined for isolated systems, with $\delta_j=0$, and $\Pi_j(\eta_j,\delta_j)$ is the same for the system out of TdE with $\delta_j>0$; $\eta_j$ is the thermodynamic equivalent of transformation, carrying information on the reaction standard change of Gibbs' free energy, initial system composition and equilibrium system composition at TdE; $\tau_j$ is the growth factor for $\delta_j$. Map (1) is nothing else but Gibbs' free energy change in open chemical system, reduced by RT; it contains a classical term $\Delta G^0_j/RT=\ln[\Pi_j(\eta_j,0)]$, corresponding to isolated state, concentration term $\ln[\Pi_j(\eta_j,\delta_j)]$, corresponding to the state with $\delta_j>0$, and a term, originating from the system interaction with its environment and reflecting the cause of the shift. Basic map derivation uses presentation of the resultant external TdF as power series of the shift, leading to a function $\varphi(\delta_j,\pi_j)$ in (1) with the system response complexity parameter $\pi_j$, equal to the highest $\delta_j$ power in the series. In our approach, at finite $\pi_j$ function $\varphi(\delta_j,\pi_j)$, splits chemical systems by two types, depending upon relationship between the external TdF and the system shift from TdE, – the strong type with

(2) $\qquad \varphi_s(\delta_j,\pi) = \delta_j(1-\delta_j^{\pi})$,

and the weak type with

(3) $\qquad \varphi_w(\delta_j,\pi) = (1-\delta_j^{\pi+1})$.

In all cases obtained expressions are logistic maps with mandatory logarithmic concentration terms. For example, basic map for the weak systems is

_________________________________

[I] System Dynamics Research Foundation, Chicago, USA, sdrf@ameritech.net.

(4) $\qquad \ln[\Pi_j(\eta_j,0)/\Pi_j(\eta_j,\delta_j)] - \tau_j(1-\delta_j^{\pi+1}) = 0.$

Graphical solutions to the maps are the diagrams of state for chemical systems, taking on shape of fork bifurcation diagrams. Plotting system deviation from TdE $\delta$ vs. $\tau$, we obtain static diagrams, plotting $\delta$ vs. TdF – dynamic diagrams. One may compare the basic types of diagrams in Fig1.

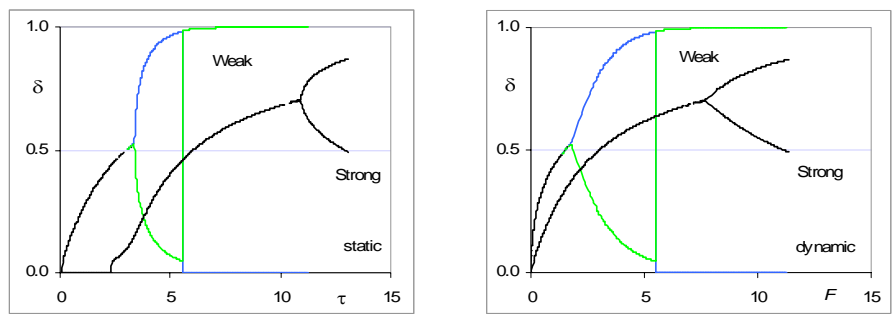

Fig1. Types of bifurcation diagrams – graphical solutions to map (4) with $\varphi_s(\delta_j,\pi)$ as (2), (2), static, and as (3), dynamic. Reaction A+B=C.

An infinite number of the diagrams of state, covering the first quadrant plane at $\delta>0$ or the third at $\delta<0$, and constituting the chemical system domain of states, may be generated by varying parameters of map (1). The bifurcation areas are the regions of the system by-stability, with or without oscillations. The latter as a well developed set of events are observed exclusively in the weak systems; appropriate static and dynamic diagrams with oscillations are compared in Fig2.

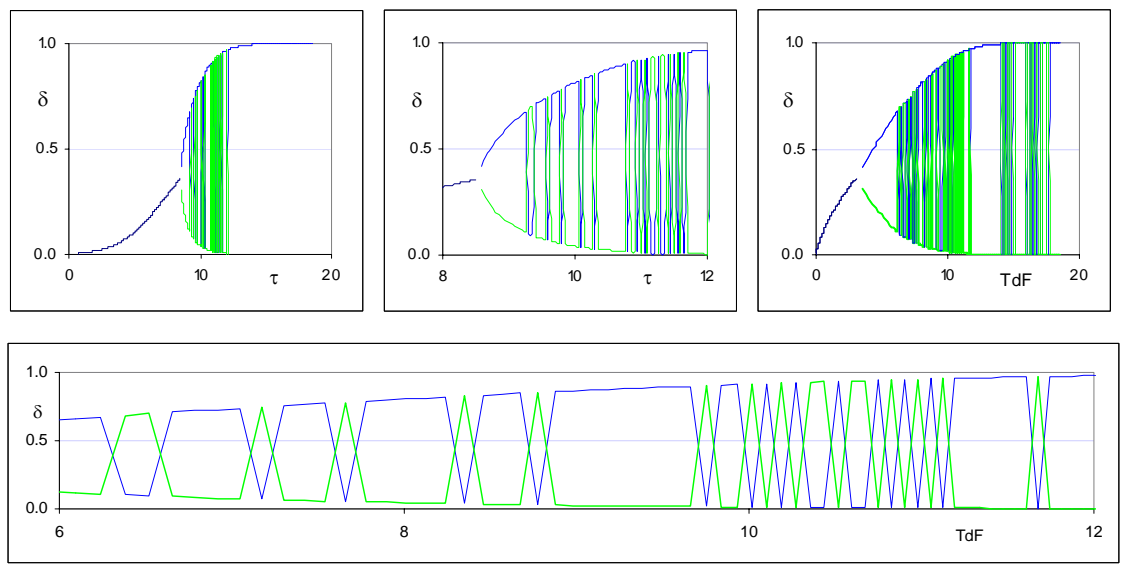

Fig2. Chemical system diagrams of states, reaction A+B=C, $\Delta G^0 = -26{,}93$ kJ/m, stoichiometric coefficients $\nu_R$, reactants, $\Sigma\nu_R = -2$, $\nu_P$, product, $\nu_P = 1$; the system complexity parameter $\pi = 1$, map (4). The upper row: weak model, static (leftmost) and dynamic (rightmost) outputs, in the middle − scaled part of the static diagram. The lower picture: a stretched part of the dynamic diagram.

DTD RESULTS FOR ELECTROCHEMICAL CELL AND LASER

In closed chemical systems the external TdF is usually of physical rather than of chemical nature. This automatically prevents material exchange between subsystems and turns the chemical subsystem into a closed entity; the electrochemical or photochemical systems are the best examples. In our research, difference of potentials on the cell electrodes was taken as the TdF in the electrochemical system. The latter comprises 2 subsystems – chemical and electrical/electrochemical, and electrochemical equilibrium means equilibrium between them. In absence of electrical current the chemical subsystem may achieve its own thermodynamic equilibrium state, but in electrochemical equilibrium it is shifted from TdE. Appropriate basic map of state for the chemical subsystem is [7]

(5) $\qquad \ln[\Pi(\eta,0)/\Pi(\eta,\delta_c)]+(1-\delta_c)nFE^{eq}/RT=0.$

with its shift $\delta_c$, the amount of transferred electron charges $n_j$, Faraday number $F$, and equilibrium cell potential $E^{eq}$. Graphical solution to (5) is shown in Fig3.

To develop DTD of lasers, we considered a quasi-chemical 2-level system with coupled excitation-irradiation reactions: up, $A+\varepsilon_p \rightarrow A^*$, $\varepsilon_p$ is the pumping energy, and down, $A^* \rightarrow A+h\nu$. Given the laser population of 1 mole (number of excitable atoms equals to Avogadro number $N_A=6.022*10^{23}$, mole$^{-1}$), in thermal equilibrium, i.e. in absence of the pumping force, the most of the laser population seats on the ground level. Its amount corresponds to the system thermodynamic equivalent of transformation for the down reaction, or $\eta_{las}$ (<1 mole). Obviously, population of the upper level is $(1-\eta_{las})$, then $\eta_{las}$ can be easily found from Boltzman-Gauss distribution

(6) $\qquad (1-\eta_{las})/\eta_{las}=\exp(\Delta G^0_{las}/RT),$

where $\Delta G^0_{las}$ is the standard energy change of the down reaction, the irradiated energy per mole of the laser excited atoms, it equals to $h\nu N_A$, and must be negative (this energy leaves the system!). The second term of map (4) is the external TdF, multiplied by $(1-\delta_{las})$, it can be recalculated into the laser pumping force in energy units. The laser states map is [8]

(7) $\qquad \lambda_\nu \nu/T + \ln p_{las} - \tau(1-\delta_{las}^{\pi+1}) =0,$

where: $\lambda_\nu=hN_A/R=47.99$, a product of the Plank, the Avogadro, and the reciprocal universal gas constants, providing the laser optical frequency $\nu$ expressed in THz; the population ratio between the upper and the ground levels in equilibrium between both processes is $p_l=[1-\eta_{las}(1-\delta_{las})]/\eta_{las}(1-\delta_{las})]$; $\delta_{las}$ – the laser shift from TdE, caused by the pumping force; $\pi$ is the laser system complexity parameter, that can be loosely defined as a number of possible ways to irradiate light via up to down transitions (e.g., in 2-level laser $\pi=1$). The $\eta_{las}$ value carries information on the energy changes in the coupled reaction at given total amount of the laser excitable atoms. One of possible graphical solutions to map (6) is shown in Fig4.

Two states of the laser coexist in bifurcation area, differing by the shifts from TdE. When the abscissa values within that area achieve certain limits, the system experiences instabilities, revealed as well pronounced chaotic oscillations. Feedback factors $(1-\delta)$ in map (5) and $(1-\delta^{\pi+1})$ with $\pi=[0,1,2,…]$ in map (6) lead to a plenty of the far-from-equilibrium events.

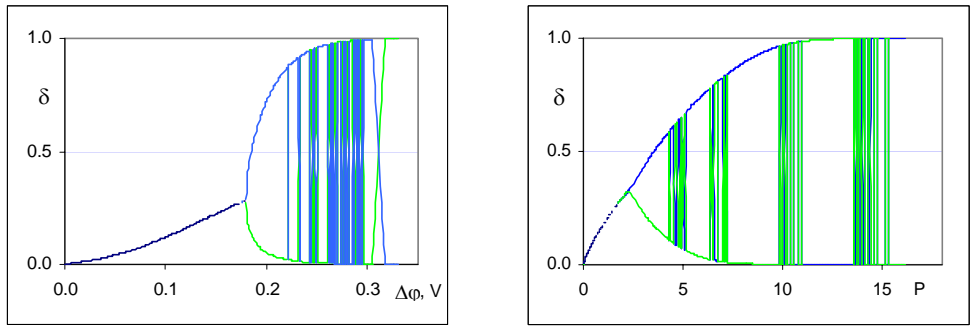

Fig.3. (left) Shift of the chemical subsystem from TdE vs. cell potential in electrochemical system, redox reaction A+B=AB, $n_j=1$, $\Delta G^0_j=-18.0$ kJ/m, T=293K, from [7].

Fig4. (right) Laser system. One of various possible dependencies of the shift vs. pumping force, $\Delta G^0_j=-94.38$ kJ/m, T=293K, $\pi=1$, from [8].

## GENERAL OSCILLATIONS IN CHEMICAL SYSTEMS, PREDICTED BY DTD

If else is not specified, in the following figures all ordinates are shifts from TdE, $\delta_j$, all abscissas are the thermodynamic forces, TdF, kJ/m, $|\Delta G^0|$ = kJ/m. One can observe strong impact of various

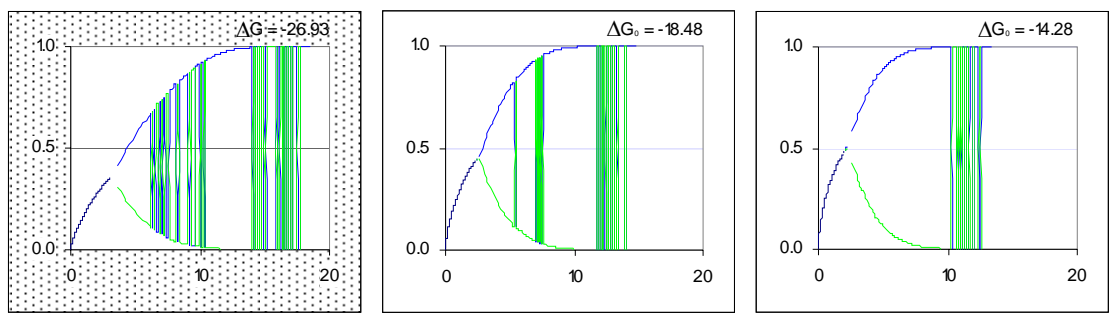

Fig5. Chemical system state diagrams, reaction aA+bB=cC ($\Sigma\nu_R$=a+b= −2, $\nu_P$=c=1, $\pi$=1). The picture with dotted frame is the comparison basis for all following series.

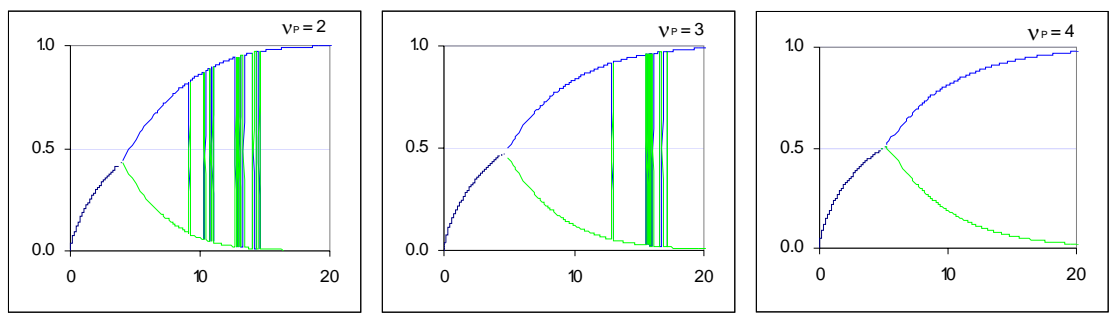

Fig6. Chemical system state diagrams, aA+bB=cC ($\Sigma\nu_R$=−2, $\pi$=1), $\nu_P$=c is varying, $\Delta G^0$= −26.93.

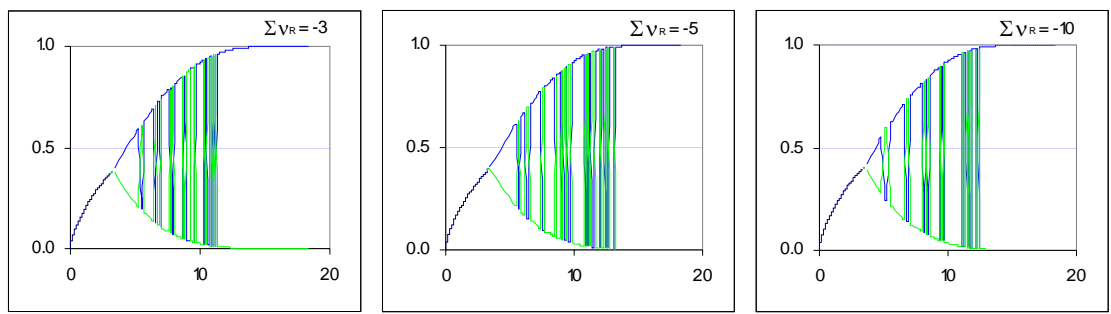

Fig7. Chemical system state diagrams, aA+bB=cC ($\nu_P$=c=1, $\pi$=1), $\Sigma\nu_R$ is varying, $\Delta G^0$ = −26.93.

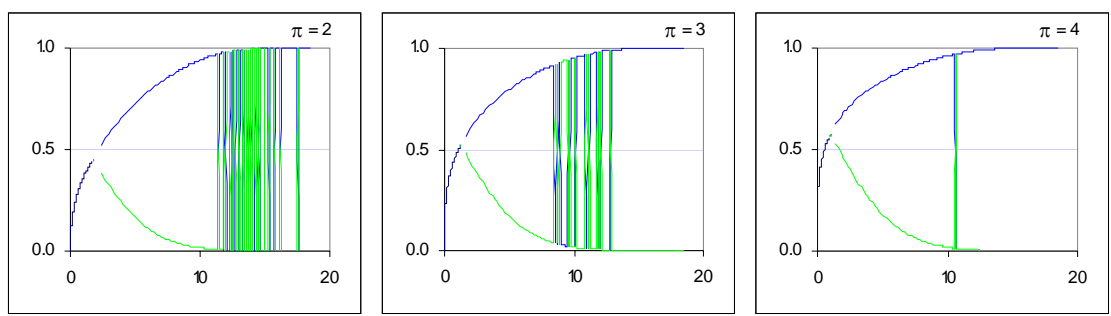

Fig8. Chemical system state diagrams, aA+bB=cC ($\Sigma\nu_R$=−2, $\nu_P$=1), $\pi$ is varying, $\Delta G^0$=−26.93.

simulation parameters on the chemical oscillations spectra, illustrated by Fig5 – Fig8. Following features, found in our observations, must be mentioned:

− spectra of chemical oscillations in closed systems depend upon such factors as $\Delta G^0$, $\Sigma \nu_R$ and the system response complexity $\pi$, but not $\Sigma \nu_P$;
− the more negative is the reaction standard change of Gibbs' free energy, the more chances are to encounter the oscillations, they are more numerous and better pronounced.

FRACTALITY OF CHEMICAL OSCILLATIONS IN CLOSED SYSTEMS

One interesting feature of chemical oscillations, found recently for electrochemical systems [7] and confirmed for some other ln-logistic maps, is fractality of the oscillation spectra, illustrated by Fig9 and Fig10. The iterative simulating program, we coded and used to investigate solutions

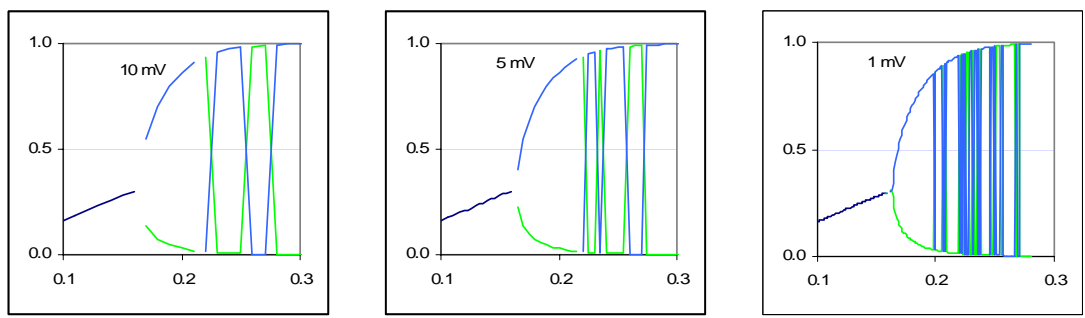

Fig.9. Fractal properties of the electrochemical oscillation spectra, $\delta_j$ (ordinate) vs. $\Delta \varphi$, (abscissa), "voltage stick" lengths are shown on the pictures, from [7].

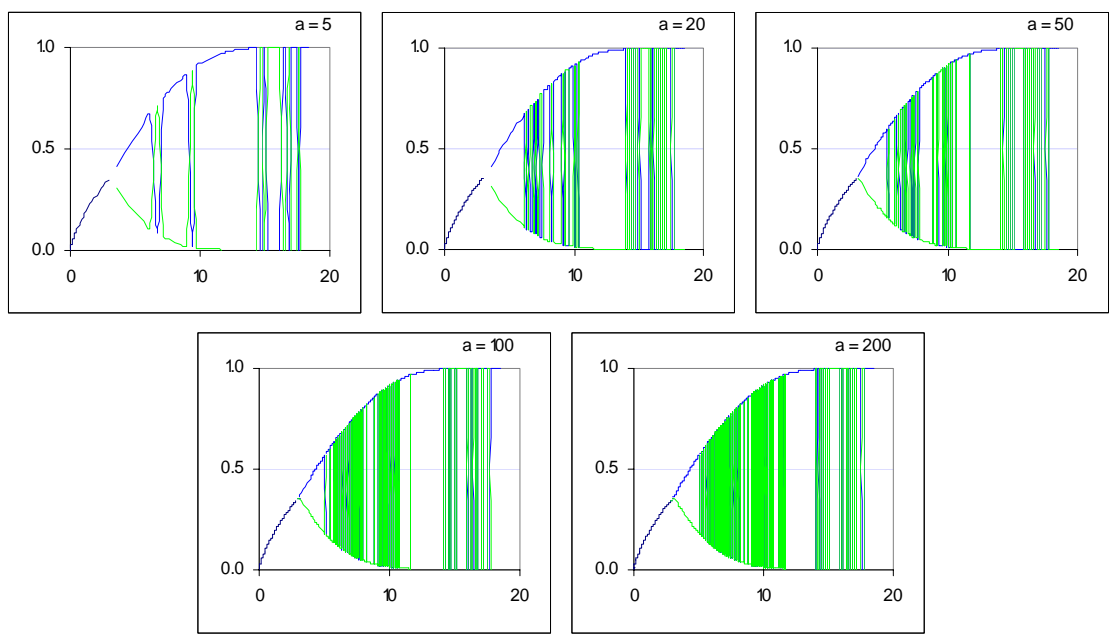

Fig10. Fractal properties of the chemical oscillation spectra, $\delta_j$ (ordinate) vs. $\tau$ (abscissa), reaction A+B=C, $\Delta G^0 = -26.93$, kJ/m, "chemical sticks" are inversely proportional to the values of a, shown on the pictures.

to ln-logistic maps, was designed to allow for scaling of major iterative parameters – it generates consecutive series of $\tau$ with set upfront variable step and for each value of $\tau$ it generates series of $\delta$ with its own independently variable and set upfront step, until appropriate map is satisfied with a given precision. Changing magnitudes of those steps means changing the "iteration" stick length and leads to fractal scaling of the oscillation spectra. Interestingly enough, bifurcation branches are not the bulk lines, but fractals by themselves, their gaps follow all oscillating ups and downs (see lower picture in Fig2, others).

CONCLUSION

We have investigated properties of the solutions to the basic DTD logistic maps, which we believe to define essentially the behavior of real closed chemical systems. To observe the DTD predicted oscillations, one has to move the system out of TdE towards its initial state by a chemical distance of $\delta_j$ or to stop the reaction running at a distance $\Delta_j=1-\delta_j$ from its starting point. Such a movement can be done in closed systems, where the above described events are driven by an external physical force. It seems to be not occasional, that the first oscillations have been detected in electrochemical systems by Fechner [10], a century before Bray had discovered oscillations in open chemical systems, driven by a competitive chemical force [11].

As concerns to the oscillations fractality, we encounter exactly the same as in the Mandelbrot's task of the Great Britain coastal line [12], where changing the measuring stick length leads to changes in the coastal length – the shorter is the stick, the more oscillations we discover. Due to obvious trend of the oscillations to group into packs it makes no sense to evaluate their fractal dimensions. Chemical oscillations fractality in some systems doesn't mean that they are fictitious – as well as fractal dimension of the coastal line of Great Britain doesn't make fictitious the shore line or the whole country. That's what the theory predicts, and the correct question would be – how to relate the scaling parameters and fractality of chemical oscillations to experimental conditions? Even any relation between fractal dimensions of the spectra and the stick length, following from the above results, brings up very little information until we know the answer to the above question. We don't know it yet.

The oscillations that we described were chaotic, having no systematic frequency or a systematic pattern. In a sense they are forced, becoming possible due to the system shifting to the far-from-equilibrium region by an external force. At the same time they are triggered not by that force immediately – at such a distance from equilibrium the system experiences huge stresses, leading to instabilities and relieving themselves through spontaneous chaotic oscillations.

One should keep in mind that the oscillations proceed when the value of external force changes; they are intrinsic for closed or open chemical subsystems. Actually the same occurs in open chemical systems, where both or even more chemical counterparts contribute to the whole system chemical oscillation, like BZ or Bray-Liebhafsky reactions. In closed chemical systems oscillations last as long as external TdF is applied, while in absence of external TdF of physical nature in open systems they fade and stop, when the active chemical reactants are exhausted.

Because all points on the bifurcation branches are in equilibrium with their antipodes, the above described oscillations, to be accomplished in real chemical systems, may demand some physical displacements within the systems. One may say with a certain caution, that thermodynamics predicts the vertical size of oscillations, while chemical kinetics may define their slope, or how they are stretched along abscissa. Quite possible, that the slopes may not be reversible with regards to the sign of the force changes, and the real pictures may be not symmetrical.

We anticipate a question – do all chemical reactions experience oscillations of the kind, predicted by DTD? The answer is not: the strong type systems don't experience a feasible oscillations, and the weak type systems start at and continue to oscillate below a certain value of reaction $\Delta G^0$ (usually at $\eta>0.5$). Unfortunately, at the time we cannot tell between the system types *a priori*. One can find enough experimental results, on qualitative level witnessing in behalf of the simulated results, presented in this paper; for a comparison of the DTD results in electrochemical systems to some experimental data we refer the reader to [7].